\documentclass{llncs}
\usepackage{dafny}
\usepackage{color}
\usepackage{graphicx}
\usepackage{mathpartir}
\usepackage{wrapfig}
\begin{document}

\newcommand{\comment}[1]{\textbf{\textcolor{red}{#1}}}
\newcommand{\todo}[1]{\comment{TO DO: #1}}

\newcommand{\dtac}{DTac}
\newcommand{\dtname}[1]{\textit{#1}}

\newcommand{\trans}{\noindent\textit{\textbf{Transformation:}} \vspace{-10pt}}
\newcommand{\pre}{\noindent\textit{\textbf{Preconditions:}}}

\newcommand{\args}{\noindent\textit{\textbf{Arguments:}}}
\newcommand{\cor}{\vspace{-10pt}\noindent\textit{\textbf{Correctness:}}}
\newcommand{\dtline}{\vspace{-5pt}\noindent\hrulefill \\}

\newcommand{\pos}[0]{\noindent\textit{\textbf{pos}}}
\newcommand{\tsymb}[1]{$\quad\Longrightarrow\quad$}
\newcommand{\rewrite}[0]{\underline{\textit{\textbf{rewrite}}}}
\newcommand{\apply}[0]{\underline{\textit{\textbf{apply}}}}
\newcommand{\pmatch}[0]{\underline{\textit{\textbf{pmatch}}}}
\newcommand{\pv}[0]{\underline{\textit{\textbf{pv}}}}
\newcommand{\einst}[0]{\underline{\textit{\textbf{inst}}}}
\newcommand{\eflush}[0]{\underline{\textit{\textbf{flush}}}}
\newcommand{\unfold}[0]{\underline{\textit{\textbf{unfold}}}}

\newcommand{\ordt}{\noindent\textit{\textbf{or}}}
\newcommand{\when}{\noindent\textit{\textbf{when}}}
\newcommand{\then}{\noindent\textit{\textbf{then}}}
\newcommand{\match}{\noindent\textit{\textbf{match}}}
\newcommand{\up}{\noindent\textit{\textbf{up}}}
\newcommand{\down}{\noindent\textit{\textbf{down}}}
\newcommand{\srcline}{\noindent\textit{\textbf{line}}}

\newcommand{\infb}[1]{\langle #1 \rangle}
\newcommand{\rulem}[0]{\Downarrow_m}
\newcommand{\ruleapp}[0]{\Downarrow_{app}}
\newcommand{\ruledt}[0]{\Downarrow_{dt}}

\lstset{language=dafny, frame=single} 
\newcommand{\scode}[1]{\lstinline[basicstyle=\normalsize\sffamily]!#1!}
\newcommand{\kw}[1]{\scode{#1}}

\title{Some Ideas for Program Verifier Tactics}
\titlerunning{Some Ideas for Program Verifier Tactics}  

\author{Gudmund Grov}
\authorrunning{Gudmund Grov} 
\tocauthor{Gudmund Grov}

\institute{
\vspace{-7pt}
School of Mathematical and Computer Sciences, \\
Heriot-Watt University, Edinburgh, UK,
\email{G.Grov@hw.ac.uk}}

\maketitle              
\pagestyle{plain}

\begin{abstract}
\vspace{-18pt}
A program verifier is a tool that can be used to verify that a ``contract"
for a program holds -- i.e. given a precondition the program guarantees that a given postcondition holds -- by only working at the level of the annotated program.
An alternative approach is to use an interactive theorem prover, which enables
users to encode common proof patterns as special programs called ``tactics". This
offers more flexibility than program verifiers, but at the expense of skills required by the user. Here, we add such flexibility to program verifiers by developing ``tactics" as a form of program refactoring called \dtac{}s. A formal characterisation and set of examples are given, illustrated with a case study  from NASA.
\end{abstract}


\vspace{-23pt}
\section{Introduction}
\vspace{-7pt}

Properties that programs should satisfy are commonly expressed by \emph{contracts} -- given a precondition the program guarantees that a given postcondition holds.
\emph{Program verifiers} can then be used to verify that a contract is satisfied
by automatically generating verification conditions (VCs), which are sent to an underlying theorem prover. Failures to prove VCs will then be highlighted in the 
text, and the user must then update the code with \emph{auxiliary annotations} to
guide the proof. Program verifiers which follow this approach 
include Spec\# \cite{Barnett04}, VCC \cite{Cohen09}, Verifast \cite{Jacobs10}, Dafny \cite{Leino10}, and the 2014 version of SPARK \cite{Barnes97}.

An alternative is to translate the annotated programs to an \emph{interactive theorem prover} (ITP) and generate and prove the VCs within this system. However, the disadvantage with this approach is that the user requires expertise in the ITP system as well as the ability to specify and implement the desired program. 
Many argue that it takes at least six months to become a  confident user of an ITP system (see \cite{Leino13a}). Such
ITP expertise is not required for program verifiers.

An ITP system keeps a stack of open goals, starting with the singleton stack containing  the  VC. Here,
the user can interactively guide the proofs, often in a backwards manner where a `proof step' involves applying a type of program called a \emph{tactic} to the 
top of the stack, and pushing any newly generated sub-goals to the stack afterwards \cite{GOR79a}. 
Finding a proof is often a trial-and-error process, which iterates until the stack is empty.
A key feature of ITP systems is that common reasoning patterns can be encoded as new tactics, which in most cases are combinations of existing ones. 
This enables automation of common tasks.

A `proof step' in a program verifier involves changing, and in most cases adding, auxiliary annotations. This author is not aware of any program verifier
with the support for encoding reasoning patterns -- and as a result
the user will need to manually encode every `proof step'. As some trial-and-error (i.e. search) may be required, such encoding can be a very 
tedious and cost-ineffective process. To enable a user to focus time and effort on key decisions, we hypothesise that:
\begin{quote}\it  
\vspace{-1pt}
It is possible to develop ``tactics" to automate program verifiers as tactics have been used to automate interactive theorem provers.
\vspace{-1pt}
\end{quote}  
In order to integrate manual proof steps with such ``tactic" applications, the
user will need work on the annotated program, meaning that a ``tactic" application will become a \emph{program transformation}. For example, a typical step is to add an intermediate \emph{assertion} in the code to guide the prover, which means the ``tactic" is a transformation which adds that assertion to the code. Other advantages
of working at the source code level are that users are not required to have additional expertise, and can inspect the changes a ``tactic" has made.

Most modern ITP systems are based on what is known as the LCF approach \cite{GOR79a}. Here, the underlying type system ensures that the only 
trusted part is a small `kernel' of axioms/rules (together with the type
system itself). Thus, if this is sound, then it is guaranteed that
(a) the proof is sound; and (b) the original conjecture to be proven is 
not changed by a tactic. In the ``tactics" proposed here a program is transformed to a new program, and the program verifier is re-applied. Thus, (a) is reduced to the correctness
of the program verifier, and we are left to focus on (b). Here, one has to 
ensure that neither the original contract nor the program is changed -- only auxiliary annotations can be altered. This can be seen as a special case of a
 behaviour preserving transformation known as  \emph{program refactoring}  \cite{Fowler99}, so we further hypothesise that:
 \begin{quote}\it
 \vspace{-1pt}
It is possible to encode  program verifier ``tactics" as a form of program \emph{refactoring}, to ensure that the contract and program are not changed.
\vspace{-1pt}
\end{quote}
To address these hypotheses we will focus on the Dafny system \cite{Leino10}, described next with an analysis of the proof process. Dafny ``tactics", called \dtac{}s, are defined, formalised and exemplified in
\S \ref{sec:dtac}, and applied to a case study from NASA in \S \ref{sec:safer}. Finally,
we conclude and discuss related and future work in \S \ref{sec:conc}.

\vspace{-5pt}
\section{Program Verification in Dafny}\label{sec:dafny}
\vspace{-7pt}

Dafny \cite{Leino10} is a programming language and program verifier for the .NET platform, developed by Microsoft Research. The language is an imperative object-oriented language, containing both methods and (proper) functions (without side-effects). We may used the term `methods' for both methods and functions
for ease of reading. Fig. \ref{fig:dafnyex} illustrates a contract for a Dafny method, where \scode{requires} precedes a precondition and \scode{ensures} precedes
a postcondition. By declaring this method as \scode{ghost}, it becomes a specification element, and will not be compiled. In fact, a ghost method can be seen as a ``lemma".
Variables can also be \scode{ghost} and all variables inside a ghost method are automatically treated as ghost variables.
Other annotations include assertions (\scode{assert}) and loop invariants (\scode{invariant}).

Note that programming language elements are often used to support the verification of ``lemmas", thus there is close correspondence here with proofs and programs.
This is exemplified in Fig. \ref{fig:dafnyex}. As a result, Dafny has been suggested beyond what is considered to be `software verification' and also as an alternative to ``traditional" theorem provers. One example is found in \cite{Leino12a}, where Dafny had comparable results to inductive theorem provers. 

\newcounter{linenumb}
\newcommand{\nl}{\stepcounter{\\
numb}{\scriptsize{\thelinenumb}}. }
\lstset{language=dafny, frame=single, numbers = left,stepnumber=1,basicstyle=\scriptsize\sffamily} 
\begin{figure}
\vspace{-30pt}
\begin{center}
\begin{minipage}{0.55\textwidth}
\indent\begin{lstlisting}
 ghost method LemmaLength4(n : int)
   requires n >= 0;
   ensures exists xs :: length(xs) == n;{
     if (n == 0){ }else{
	   LemmaLength4(n-1);
	   var xs :| length(xs) == n-1; }}
\end{lstlisting} 
\end{minipage}
\end{center}
\vspace{-10pt}
\vspace{-10pt}
\caption{(Almost) a proof of a ``lemma" in Dafny (adapted from \cite{Leino13a}).} \label{fig:dafnyex}
\vspace{-16pt}
\end{figure}
\lstset{language=dafny, frame=none, numbers=none,basicstyle=\footnotesize\sffamily} 

Fig. \ref{fig:dafnyex} represents a ``lemma" which states that for a given non-negative number \scode{n}, there exists a list of that length. The proof is by induction.
For the step case (\scode{n} $\neq$ \scode{0}) the induction hypothesis is applied by making a recursive call to itself with \scode{n-1} as an argument. This means that we can assume the 
postcondition of the recursive call, i.e. \scode{exists xs. length(xs) == n-1}. Using this property, line 5 creates such a list
 (\scode{:}\scode{|} should be read ``such that"). No guidance is given for the base case (\scode{n==0}) as this is trivial to prove for Dafny.
  
To verify the program, Dafny translates it into an \emph{intermediate verification 
language} (IVL), called Boogie2 \cite{Barnett06}. An IVL can be seen as a layer to ease the process of generating new program verifiers. From Boogie2 a set of VCs are generated and sent to the Z3 SMT solver \cite{Moura08}. If it fails, then the failure is translated back to the Dafny code, via Boogie2.
To illustrate, the \scode{LemmaLength} method above does not actually verify. The problem is that Dafny is not able to determine a 
witness for the existential in the postcondition. Here, the user needs to give Dafny a hint, achieved by adding the following assertion after line 5:
\vspace{-3pt}
\begin{lstlisting}
  assert length(Cons(1,xs)) == n;
\end{lstlisting}
\vspace{-3pt}
Moreover, since \scode{LemmaLength} is marked to be a \kw{ghost method}, it is only there to aid the verification process.  Thus, coming up with this in the first place is likely to have been part of the proof 
process in \cite{Leino13a}. Such ``proof steps" in Dafny, will form the basis for 
a ``tactic" in this context. Such steps include:
\begin{description}
\item[Add assertions.] A simple ``lemma" can be represented as an assertion in the text,
as illustrated above with \scode{assert length(Cons(1,xs)) == n}.
\item[Create ghost methods.] More involved ``lemmas", such as \scode{LemmaLength} itself that require further proofs,
can be captured by a ghost method, where the ``lemma" is the postcondition and any required conditions are preconditions. 
\item[Add preconditions.] In certain cases one can restrict the applicability of methods with preconditions. 
To illustrate, when creating \scode{LemmaLength}, the \scode{n >= 0} precondition could have been added in a second step. 
\item[Proof by cases.] One may need to add code in a ghost method to guide a proof. One common step is to introduce cases, as shown 
on lines $3-5$ of \scode{LemmaLength} by using an \scode{if-else}  
statement.
\item[Add ghost method calls.] One may need to make calls to ghost methods in order
to apply ``lemmas". This was illustrated for \scode{LemmaLength} by making a call to itself (to apply the induction hypothesis).
\item[Add ghost variables.] This is illustrated on line 5 for the \scode{LemmaLength} method (variables declared in
ghost methods are implicitly ghost variables).
\item[Add postcondition.] One illustration of this is in terms of recursion. A more general lemma is often easier to prove, thus, by strengthening the postcondition the proof may become simpler.
\end{description}
We also consider loop invariant generation an important feature. However, we will not discuss
that further here.
Our goal is to be able to encode these steps as ``tactics" to automate 
common proof steps for program verifiers.

\vspace{-5pt}
\section{\dtac{}s -- \underline{D}afny \underline{Tac}tics } \label{sec:dtac}
\vspace{-7pt}

To illustrate how an LCF tactic works with the stack of sub-goals, consider the
stack [$A \wedge B$,$C$], and the well-known conjunction introduction (\textit{conj-I}) tactic:
$\frac{\vdash A \qquad \vdash B}{\vdash A \wedge B}$. Applying \textit{conj-I}
will reduce the first sub-goal $A \wedge B$ into the two sub-goals [$A$,$B$], which
are popped to the stack: [$A$,$B$,$C$]. 

In a program verifier such as Dafny, failures to prove verification conditions are highlighted in the source code, thus an `open goal' should here be seen as:
the type of failure (e.g. a postcondition), properties of the failure (e.g which postcondition), and the position in the code. The sum of these open goals then correspond to the stack in an LCF prover. A Dafny ``tactic", which we call \dtac{}, is  a code transformation which creates new (ideally easier to prove) failures when reapplying the program verifier. To illustrate, consider this \emph{MainGoal()} method:
\begin{lstlisting}
   method MainGoal() ensures A && B { ... }
\end{lstlisting}
\noindent Assume that the postcondition fails to prove. Here, a \dtac{} which corresponds to the \emph{conj-I} tactic will, when applied, 
create one \scode{ghost method} for each conjunct and make a call to each of them within the \kw{MainGoal()} method:
\begin{lstlisting}
   ghost method SubGoalA() ensures A { ... }
   ghost method SubGoalB() ensures B { ... }
   method MainGoal() ensures A && B { SubGoalA(); SubGoalB(); }
\end{lstlisting}
The program verifier will then solve \kw{MainGoal()}, but \kw{SubGoalA()}
and \kw{SubGoalB()} may fail to verify, creating one or two new sub-goals. Thus, the original goal has been reduced into simpler goal(s).

While the LCF approach handles soundness by reducing every proof to the
``kernel" of trusted rules \cite{GOR79a}, we need to restrict the type of transformation allowed, and start with the concept of 
\emph{program refactoring}. i.e. \emph{``a program transformation which preserves the external behaviour of the original program"} \cite{Fowler99}. However, the motivation behind our transformations is to prove that a program satisfies a contract, thus the program should not change at all. 
Moreover, we cannot allow arbitrary changes to the annotations and contracts. However, as the \emph{conj-I} example showed, a proof is still conducted by changes to the annotations.  
Thus, we constrain which parts of the code can be changed and which are ``final",
and define a \dtac{} as follows:
\begin{definition} \label{def:dtac}
A method is either marked as `\emph{private}' or `\emph{public}' -- or it has been `\emph{generated}' by a \dtac{}. We separate between the `\emph{code}' and `\emph{specification}' of a program, where the latter is any elements that are not compiled, i.e. contracts, annotations and \kw{ghost} code.
A \dtac{} is then a refactoring which 
\vspace{-7pt}
\begin{itemize}
\item cannot change the code for `\emph{private}' and `\emph{public}' methods;
\item cannot strengthen preconditions or weaken postconditions for
 `public' methods, but make arbitrary changes to all other specification elements. 
\end{itemize}
\end{definition}
We are interested in proving that the contracts of `public' methods hold --
a `private' method is simply there as a helper method, and we can change the contract as seen fit. The intuition behind allowing this is that if we e.g. strengthen a precondition, then the caller must prove that it is satisfied when calling it. Anything `generated' should be seen as part of the proof process and can thus be changed arbitrarily. Note that a user needs to declare which elements are `private' and which are `public'. From definition \ref{def:dtac} we can see that:
\begin{theorem}
A \dtac{} does not change the program and preserves the contract of `public' methods.
\end{theorem}
\vspace{-12pt}
\begin{proof}
This follows directly from Definition \ref{def:dtac}.
\end{proof}
\lstset{language=dafny, numbers=none, frame=single,basicstyle=\scriptsize\sffamily}
The following BNF gives the grammar for  \dtac{}s:

\dtline
\vspace{-20pt}
\begin{tabbing}
$\quad$\=$\langle$dtac$\rangle\;\;$ \= := $\;\;$\= $\langle$name$\rangle$`(' $\langle$arg$\rangle^*$ `)' `$:=$' $\langle$body$\rangle$ \\ 
\> $\langle$body$\rangle$ \> $:=$ \> \when{} $\langle$prop$\rangle$ \then{} $\langle$trans$\rangle$ $\quad | \quad$ $\langle$trans$\rangle$ \\
\> $\langle$trans$\rangle$ \> $:=$ \>  $\langle$code$\rangle$  
$\Longrightarrow$ $\langle$code$\rangle$  [ $\langle$inst$\rangle$  ]$\;\; | \;\;$  \match{} $\langle$code$\rangle$  [ $\langle$inst$\rangle$  ]  \\
\> \> $|$ \>  $\langle$trans$\rangle$ `;' $\langle$trans$\rangle$  $\;\; | \;\;$  $\langle$name$\rangle$`(' $\langle$arg$\rangle^*$ `)' [ $\langle$inst$\rangle$  ] \\
\> \> $|$ \>  \ordt{}($\langle$trans$\rangle$,$\langle$trans$\rangle$) \\
\> $\langle$inst$\rangle$ \> $:=$ `['($\langle$pos$\rangle$ $|$ `$?$'$\langle$name$\rangle$ `:=$$'$\langle$code$\rangle$)$^*$`]` \\
\> $\langle$pos$\rangle$ \> $:=$ \> `@'$\langle$name$\rangle$ $\; | \;$ \srcline{}`('$\langle$nat$\rangle^*$`)'$\; | \;$ \up{}`('$\langle$pos$\rangle$`)' $\;\; | \;\;$ \down{}`('$\langle$pos$\rangle$`)'
\end{tabbing}
\vspace{-7pt}
\dtline
`${*}$' represents repetition, `$[-]$' for optional, and `$|$' for alternatives.
$\langle$code$\rangle$ is the code and annotation of Dafny, populated with variables (preceded by `?') and a special \rewrite{} function. This
function is overloaded to work for: a single rewrite rule $\rewrite (l \rightarrow r,c)$;  a triple $\rewrite (l,r,c)$ or a list $\rewrite ([l_1,\cdots,l_n],$ $[r_1,\cdots,r_n],c)$, which is applied pairwise  $\rewrite (l_i,r_i,c)$. In all cases
the code $c$ is rewritten. We refer to standard rewriting literature for details \cite{Baader98}. For readability, we will also use ellipses `\scode{...}' for parts of the code that can be ignored instead of variables.  $\langle code_1\rangle$  
$\Longrightarrow$ $\langle code_2\rangle$ represents a rewrite rule, and thus $\langle code_2\rangle$ cannot contain any variables that are not present in $\langle code_1\rangle$. `;' is sequential composition, while \when{} and \then{}
are used to express conditions. \match{} is used to check for particular code patterns and instantiate variables, typically used with a sequentially composed \dtac{}s. \ordt{} is used to express choice. 

We assume the presence of an \emph{environment} E, which contains binding of variables and positions which are preceeded by `@'. $\langle$prop$\rangle$ is a predicate on the underlying environment and the source code. In order to have a simpler model of the composition of \dtac{}s, the variables bindings from the first are
removed before applying the second \dtac{}. This introduces a form of ``locality" of the namespaces: e.g. for $dt_1;dt_2$, if both binds say variable $?x$ then these can be seen as separate variables. To carry such bindings between \dtac{}s an instantiation environment $\langle inst \rangle$ can be provided. Here, variables can be bound by $[?x := e]$, where $e$ uses the existing environment.
To illustrate in $dt_1;dt_2[?x := ?x]$, $?x$ in $dt_2$ is defined to be the same $?x$ as in $dt_1$.
A second usage of the instantiation environment is to constrain the position in the source code 
where a \dtac{} can be applied -- either by giving reference to special comment preceded by `@', or
line numbers. If $dt[@p1,@p2]$, then $dt$ can only be applied at one of these positions.
Inspired by Huet's \emph{Zippers} \cite{Huet97}, we can also move \up{} and \down{} a statement relative to a given position. 
Note that \up{}/\down{} are only applicable `within' a method -- they will fail if attempting to e.g. move down when at the
last statement of the method. Positions are very useful when composing \dtac{}s of the method. Recursive application of \dtac{}s is prohibited.
Fig. \ref{fig:dtacsex} contains 
several example \dtac{}s. Given source code $c$ and a \dtac{} $dtac$, $\Downarrow$ refactors $c$ into $c'$:
\begin{small}
$$
 \inferrule
 {
 E \in \pv(c) \\ \infb{E,c,dtac} \ruledt \infb{\_,c'} \\ 
 \vdash c'
 }
 {
   \infb{c,dtac} \Downarrow c'
 }
$$ 
\end{small}
\noindent \pv(c) can be seen as the verification condition generator, and returns a set of 
environments, each containing the bindings: $?error$ for the type of error;  $?err\_arg$ for the property that failed; and \emph{@err\_pos} for the position of the error. $\ruledt$ then evaluates the \dtac{} and code under this environment.
The new source has to be type correct, and we assume the presence of a typing relation $\vdash$.
 
\begin{figure}[h]
\vspace{-7pt}
\dtline{}
\vspace{-8pt}
$$
\inferrule[seq]
 {
   \infb{E,c,dt_1} \ruledt \infb{E'',c''} \\\\    \infb{E'',c'',dt_2} \ruledt \infb{E',c'} 
 }{
   \infb{E,c,dt_1 ; dt_2} \ruledt \infb{E',c'}
 }
 \qquad
  \inferrule[dtac]
 {
  dtac \notin \{\textbf{or}(\_,\_),(\_~;~\_)\} \\\\
  \langle{} E,c,dtac \rangle{} \rulem E'' \\
  \langle{} E'',c,dtac \rangle{} \ruleapp \infb{E',c'} 
  }
  {
   \langle{} E,c,dtac \rangle{} \ruledt \infb{E',c'}
 } 
 $$ 
 $$
 \inferrule[or1]
 {
   \infb{E,c,dt_1} \ruledt \infb{E',c'} 
 }{
   \infb{E,c,\ordt(dt_1,dt_2)} \ruledt \infb{E',c'}
 } 
 \quad
 \inferrule[or2]
 {
   \infb{E,c,dt_2} \ruledt \infb{E',c'} 
 }{
   \infb{E,c,\ordt(dt_1,dt_2)} \ruledt \infb{E',c'}
 } 
$$
 $$
  \inferrule[match1]
 {
  \infb{E,c,c} \rulem E' \\ \infb{E',c} \models p
  }
  {
   \langle{} E,c,\when{}~p ~\then{}~ c \rangle{} \rulem E'
 } 
 \qquad
  \inferrule[match2]
 {
  E' \in \pmatch(\eflush(E) \cup \einst(inst,E),c,c_1)
  }
  {
   \infb{E,c,c_1 \Longrightarrow c_2~inst} \rulem E'
 } 
$$
$$
  \inferrule[match3]
 {
  E' \in \pmatch(\eflush(E) \cup \einst(inst,E),c,m)
  }
  {
   \infb{E,c,\match{}~m~inst} \rulem E'
 } 
 \quad
   \inferrule[match4]
 {
  \infb{E,c,\unfold(dt(args))~inst} \rulem E'
  }
  {
   \infb{E,c,dt(args)~inst} \rulem E'
 } 
 $$
  $$
  \inferrule[app1]
 {
  c' \in \apply(E,c_1 \rightarrow c_2,c) \\\\
  E' = E[@pos,@start,@end := @m,@s,@e]
  }
  {
   \infb{E,c,c_1 \Longrightarrow c_2~inst} \ruleapp \infb{E',c}
 } 
\;
  \inferrule[app2]
 {
      E' = E[@pos,@start,@end := \\\\  @m,@s,@e]
  }
  {
   \infb{E,c,\match{}~m~inst} \ruleapp \infb{E',c}
 } 
 $$
 $$
   \inferrule[app3]
 {
  \infb{E,c,\unfold(dt(args))~inst} \ruleapp \infb{E',c'}
  }
  {
   \infb{E,c,dt(args)~inst} \ruleapp \infb{E',c'}
 } 
\quad
   \inferrule[app3]
 {
  \infb{E,c,c} \ruleapp \infb{E',c'}
  }
  {
   \infb{E,c,\when{}~p ~\then{}~ c \rangle{}} \ruleapp \infb{E',c'}
 } 
 $$ \\

\vspace{-18pt}

\dtline{}
\vspace{-20pt} 
\caption{Evaluation semantics for \dtac{}s.} \label{fig:semantics}
\vspace{-10pt}
\end{figure}

Fig. \ref{fig:semantics} defines the $\ruledt$ relation. First sequences, and \ordt{} have the obvious semantics. Other \dtac{}s have two phases: a \emph{matching} phase followed by an \emph{application} phase.

During the \emph{matching phase}, if the \dtac{} is conditional (\when{}-\then{}-), the body is matched then the precondition is checked (by $\models$ 
whose definition is omitted). \pmatch{} takes an environment, code and a code pattern, and returns a set where each element is an environment containing 
updated with all the matches. In addition to the variables the bindings from the pattern,
`?pre' is bound to the preconditions, `?post' to the postconditions, `?meth' to the name of the method the code is within, and `?arg' to the argument of the method.
If there are e.g. many preconditions, then it is used non-deterministically to refer to any of them.
Moreover, `@s',`@e' and`@m' become bound to the full, start and end positions.
$\einst(inst,E)$ instantiates the variables in $inst$ using $E$, while 
$\eflush$ removes all binding created by a pattern (e.g. 
`?pre', `@pos', and `?error' are not removed). \unfold{} unfolds the given \dtac{} and for simplicity we assume that it has access to all of the definitions. 

In the \emph{application phase} $\apply{}(E,r,c)$ represents the application of a rule $r$ under the environment $E$ to the code $c$. Here, `@m' is used to apply it 
to the same position as the match from the matching phase. For \pmatch{} and \apply{} we refer to standard rewriting literature, such as \cite{Baader98}, and omit the detail.

From $\Downarrow$ we can see that the composition of \dtac{}s preserves the \dtac{} validity;

\begin{theorem}
If $dt_1$ and $dt_2$ are \dtac{}s then $dt_1~;~dt_2$ and $\ordt(dt_1,dt_2)$ are \dtac{}s.
\end{theorem}
\vspace{-10pt}
\begin{proof}
$\ordt(dt_1,dt_2)$ follows directly from $\ruledt$ and the assumption as either $dt_1$ or $dt_2$ is applied. 
Following $\ruledt$, $dt_1~;~dt_2$ first applies $dt_1$, giving $c''$, which by the assumption has only
made valid changes following Definition \ref{def:dtac}. $dt_2$ is then applied to $c''$, which by the assumption
has only made changes following Definition \ref{def:dtac}. Thus, the $dt_1~;~dt_2$ is a \dtac{} following
 Definition \ref{def:dtac}. 
\end{proof}



\begin{figure}
\vspace{-10pt}

\dtline 
\vspace{-18pt}
\begin{small}
\begin{tabbing}
\=\dtname{assert-I}(P) $\quad\;\;$\= $:=  \quad$ \=
\begin{minipage}{0.03\textwidth}
\begin{lstlisting}

\end{lstlisting}
\end{minipage}
\tsymb{\dtname{assert-I}}
\begin{minipage}{0.3\textwidth}
\begin{lstlisting}
assert ?P; /*@ass*/
\end{lstlisting}
\end{minipage} \\[-7pt]
\>\dtname{post-I}(P) \> $:=$ \>
\begin{minipage}{0.3\textwidth}
\begin{lstlisting}
method ?m(..) ... { ... }
\end{lstlisting}
\end{minipage}
\tsymb{\dtname{post-I}}
\begin{minipage}{0.25\textwidth}
\begin{lstlisting}
method ?m(..) ...
  ensures ?P { ... }
\end{lstlisting}
\end{minipage} \\[-7pt]
\>\dtname{pre-I}(P) \> $:=$ \> \when{} $?m$ is not public \\[-5pt]
\>\>\> \then{} $\quad$
\begin{minipage}{0.2\textwidth}
\begin{lstlisting}
method ?m(..) ... 
\end{lstlisting}
\end{minipage}
\tsymb{\dtname{pre-I}}
\begin{minipage}{0.25\textwidth}
\begin{lstlisting}
method ?m(..) 
  requires ?P ...
\end{lstlisting}
\end{minipage} \\[-7pt]
\>\dtname{assert-E}() \> $:=$ \>
\begin{minipage}{0.15\textwidth}
\begin{lstlisting}
assert P;
\end{lstlisting}
\end{minipage}
\tsymb{\dtname{assert-E}}
\begin{minipage}{0.18\textwidth}
\begin{lstlisting}
/* @ass */
\end{lstlisting}
\end{minipage} \\[-7pt]
\> \noindent\dtname{pre-E}() \> $:=$ \>
\begin{minipage}{0.35\textwidth}
\begin{lstlisting}
method ?m(..) requires ?P ...
\end{lstlisting}
\end{minipage}
\tsymb{\dtname{pre-E}}
\begin{minipage}{0.2\textwidth}
\begin{lstlisting}
method ?m(..) ... 
\end{lstlisting}
\end{minipage} \\[-5pt]
\> \dtname{post-E}() \> $:=$ \> \when{} $?m$ is not public \\
\> \>\> \then{} $\quad$
\begin{minipage}{0.25\textwidth}
\begin{lstlisting}
method ?m(..) ... 
  ensures ?P { ... }
\end{lstlisting}
\end{minipage}
\tsymb{\dtname{post-E}}
\begin{minipage}{0.22\textwidth}
\begin{lstlisting}
method ?m(..) ... { ... }
\end{lstlisting}
\end{minipage} \\[-7pt]
\> \dtname{post-to-assert}() \> := \> \match{} ~\begin{minipage}{0.22\textwidth}
\begin{lstlisting}
method ?m(..) ... 
 ensures ?P ...
\end{lstlisting}
\end{minipage} ; \dtname{assert-I}(?P)[@end,?meth := ?m] \\[-5pt]
\> \dtname{assert-to-pre}() \> := \>
 \dtname{assert-E}()[@start];~\dtname{pre-I}(?P)[?meth := ?meth] \\[-1pt]
\> \dtname{assert-to-post}() \> := \>
\match{} ~\begin{minipage}{0.14\textwidth}
\begin{lstlisting}
?m(?xs);
assert ?P;
\end{lstlisting}
\end{minipage}~; ~\dtname{assert-E}()[@pos,?P := ?P,?m := ?m]~;~ \\[-5pt]
\>\>\>
\match{} ~\begin{minipage}{0.18\textwidth}
\begin{lstlisting}
method ?m(?ys)
\end{lstlisting}
\end{minipage}~[?P := ?P,?m := ?m,?xs := ?xs]; \\[-5pt]
\>\>\>
\dtname{post-I}(\rewrite($?xs,?ys,?P$))[?m := ?m] \\
\> \dtname{assert-rewr}(R) \> := \> \dtname{assert-E}();\dtname{assert-I}(\rewrite(R,?P)[@pos]\\
\> \dtname{assert-up1}() \> := \> \emph{assert-E}();\emph{assert-I}(?P)[\up{}(@pos)] \\
\> \dtname{assert-up2}() \> := \>
\match{} ~\begin{minipage}{0.28\textwidth}
\begin{lstlisting}
?x := ?e; assert ?P;
\end{lstlisting}
\end{minipage} ~;~ 
\emph{assert-E}()[@pos,?P:=?P, \\[-3pt]
\>\>\>?x:=?x,?e:=?e];\emph{assert-I}(\rewrite(?x,?e,?P)[\up{}(@pos)] \\[-3pt]
\>\dtname{assert-up3}() \> $:=$ \>
\begin{minipage}{0.22\textwidth}
\begin{lstlisting}
if(..){..}else{..}
assert ?P;
\end{lstlisting}
\end{minipage}
\tsymb{\dtname{assert-E}}
\begin{minipage}{0.38\textwidth}
\begin{lstlisting}
if(..){../*@ass1/*assert ?P;}
else {../*@ass1/*assert ?P;}
\end{lstlisting}
\end{minipage} \\[-6pt]
\> \dtname{assert-up}() \> := \>\ordt(\dtname{assert-up3}(),\ordt(\dtname{assert-up2()},\dtname{assert-up1}())) \\
\> \dtname{post-to-post}() \> := \> 
\match{} ~\begin{minipage}{0.62\textwidth}
\begin{lstlisting}
method ?m1(...) ... ensures ?P; { ... ?m2(?xs); ... }
\end{lstlisting}
\end{minipage} ~;~ \\[-7pt]
\>\>\> \match{} ~\begin{minipage}{0.2\textwidth}
\begin{lstlisting}
method ?m2(?ys)
\end{lstlisting}
\end{minipage}~~[?m1:=?m1,?m2 := ?m2, ?xs := ?xs, \\[-5pt]
\>\>\> ?P := ?P] ~;~ \dtname{post-I}(\rewrite($?xs,?ys,?P$))[?meth := ?m2] \\

\> \dtname{pre-to-assert}() \> := \>  
 \when{} ?error = ``A precondition for this call might not hold" \\
\>\>\> \then{} \dtname{assert-I}($?err\_arg$)[\up(@err\_pos)] \\

\> \dtname{null-to-assert}() \> := \> \when{} ?error = ``target object may be null" \\
\>\>\> \then{} \dtname{assert-I}($?err\_arg \neq$ \scode{null})[\up(@err\_pos)] \\

\> \dtname{pred-var-I}(v,P) \> := \>  
\begin{minipage}{0.03\textwidth}
\begin{lstlisting}

 \end{lstlisting}
\end{minipage}
\tsymb{pred-var-I}
\begin{minipage}{0.35\textwidth}
\begin{lstlisting}
ghost var v :| P /* @gv */
 \end{lstlisting}
\end{minipage} \\
\> \dtname{ex-E}(x,P) \> := \> \when{} $P$ = $\exists ~?y. ?P'$ 
\then{}~ \dtname{pred-var-I}(x,\rewrite(?y,x,$?P'$)) \\
\> \dtname{case-I}(cond) \> :=\> \when{} ?meth is generated \\[-5pt]
\>\>\>\then{} $~$
\begin{minipage}{0.03\textwidth}
\begin{lstlisting}
 
 \end{lstlisting}
\end{minipage}
\tsymb{case-I}
\begin{minipage}{0.5\textwidth}
\begin{lstlisting}
if (cond){ /*@case1*/}else{/*@case2*/ }
 \end{lstlisting}
\end{minipage} \\[-7pt]
\>\dtname{call-I}(m,args)\> := \> \when{} $m$ is ghost \then{}~ 
\begin{minipage}{0.03\textwidth}
\begin{lstlisting}
 
 \end{lstlisting}
\end{minipage}
\tsymb{call-I}
\begin{minipage}{0.28\textwidth}
\begin{lstlisting}
m(args); /* @call */
 \end{lstlisting}
 \end{minipage} \\[-5pt]
\>\dtname{IH-I}()\> := \> \dtname{call-I}(?meth,?arg - 1)[@case2]
\end{tabbing} \vspace{5pt} 

\cor{} \dtname{assert-I} and \dtname{assert-up3} only adds a specification element.
\dtname{post-I} only strengthens the postcondition, while 
\dtname{pre-E}  only weakens a precondition.
\dtname{pre-I} and \dtname{post-E} have preconditions ruling out application to public methods. \dtname{assert-E} only removes an assertion
\dtname{pred-var-I} only introduces ghost code, while \dtname{case-I} and \dtname{call-I} are restricted to ghost code. The remaining are just 
composition or special cases of other \dtac{}s and are thus correct.  \\
\end{small}\vspace{-12pt}

\dtline
\vspace{-18pt}
\caption{\dtac{} examples} \label{fig:dtacsex}
\end{figure}

The first six \dtac{}s of Fig. \ref{fig:dtacsex} are adaptations of standard
introduction and elimination tactics. Note that $\Downarrow$ ensures that 
the resulting code is well-typed by $\vdash$.
\dtname{post-to-assert} turns a postcondition into an assertion at the 
end of the method position (achieved by @end), and is normally applied
due to a failed postcondition.
The \dtname{assert-to-pre} \dtac{} turns an assertion in the beginning of a method into a precondition,
and  can be seen as a way of moving the work of proving $?P$ to the caller. Similarly, \dtname{assert-to-post} moves an 
assertion to a postcondition of the method preceding it.
Here, \rewrite{} is used to adapt the arguments to the new context. E.g. if the assertion was \scode{x+1 < y} with method call \scode{?m(x+1,y)} for \scode{method ?m(a,b)}, then the postcondition would become \scode{a < b}.
\dtname{assert-rewr} applies a given rewrite rule to an assertion.
\dtname{assert-up} moves an assertion upwards in the code passing a statement. It is moved into both branches of an \scode{if-else} statement, and updated with the expression in case of an assignment (both achieved by \ordt{}). 
A failed postcondition of a method could be due to a failed postcondition of a nested method call. In this
case, the postcondition needs to be ``copied" to the method called, achieved by \dtname{post-to-post}.
\dtname{pre-to-assert} and \dtname{null-to-assert} are examples of \dtac{}s triggered by a particular type of failure, and in both cases an assertion is added just before the failure.
Given a 
fresh variable and a predicate, \dtname{pred-var-I} is used to introduce a new ghost variable, such that
the predicate holds. A special use of this is to eliminate an existential quantifier, as shown in 
the \dtname{ex-E} \dtac{}. \dtname{case-I} introduces a case split, using an \scode{if-else} statement,
whilst \dtname{call-I} inserts a call to a ghost method. \dtname{IH-I} is a special case of \dtname{call-I},
where a recursive call is made to itself, with the argument decremented by one, at the @case2 position. In addition to those in Fig. \ref{fig:dtacsex}, five more \dtac{}s are required for the case study. The definitions of these have been omitted for space reasons. \dtname{assert-down} moves an assertion down. \dtname{assert-conj-I} splits a conjunction in an assertion into one assertion for each conjunct. 
\dtname{assert-up-ctxt()} behaves as \dtname{assert-up}, but preserves
``context" when within an \scode{if(-else)} block. E.g. 
assertion \scode{P} and condition 
\scode{C}, becomes \scode{C ==> P} when moving out of the \scode{if} block and 
$\lnot$\scode{C ==> P} when moving out of the \scode{else} branch.
\dtname{assert-strengthen()} removes a condition (conjunct) from a conditional assertion. Finally, \dtname{assert-comb1()} is a very low-level \dtac{} which turns \scode{assert P ==> Q; assert P ==> R} into \scode{assert Q ==> R}.

\begin{figure}
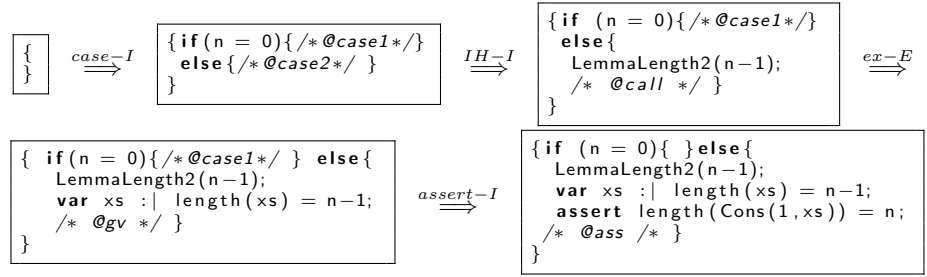

\vspace{-20pt}
\begin{center}
\begin{minipage}{0.02\textwidth}
\begin{lstlisting}
{
}
\end{lstlisting}
\end{minipage}
$~~\stackrel{case-I}{\Longrightarrow}~~$
\begin{minipage}{0.29\textwidth}
\begin{lstlisting}
{if(n == 0){/*@case1*/}
 else{/*@case2*/ }
}
\end{lstlisting}
\end{minipage}
$~~\stackrel{IH-I}{\Longrightarrow}~~$
\begin{minipage}{0.31\textwidth}
\begin{lstlisting}
{if (n == 0){/*@case1*/}
 else{ 
  LemmaLength2(n-1);
  /* @call */ }
}
\end{lstlisting}
\end{minipage}
$~~\stackrel{ex-E}{\Longrightarrow}$

\vspace{-8pt}
\begin{minipage}{0.40\textwidth}
\begin{lstlisting}
{ if(n == 0){/*@case1*/ } else{
   LemmaLength2(n-1);
   var xs :| length(xs) == n-1;
   /* @gv */ }
}
\end{lstlisting}
\end{minipage}
$~~\stackrel{assert-I}{\Longrightarrow}~~$
\begin{minipage}{0.42\textwidth}
\begin{lstlisting}
{if (n == 0){ }else{
  LemmaLength2(n-1);
  var xs :| length(xs) == n-1;
  assert length(Cons(1,xs)) == n;
 /* @ass /* }
}
\end{lstlisting}
\end{minipage}
\end{center}
\vspace{-20pt}
\caption{Proof of the \scode{LemmaLength} method using \dtac{}s.} \label{fig:lemlength}
\vspace{-15pt}
\end{figure}

Fig. \ref{fig:lemlength} is a proof of the \scode{LemmaLength ghost method}.
The starting point is the method of Fig. \ref{fig:dafnyex}, but with an empty body.
This fails to prove and the first step applies \dtname{case-I}()[?meth := \scode{LemmaLength}]. \dtname{IH-I} is then used to apply the ``induction hypothesis" where the existential is instantiated 
by \textit{ex-E}(\scode{xs},\scode{exists xs. length(xs) == n})[@call].
The proof is completed by giving the witness for the postcondition as an assertion: \textit{assert-I}(\scode{length(Cons(1,xs)) == n})[@gv]. The
next section contains a larger case study and more realistic applications of \dtac{}s.

\vspace{-5pt}
\section{A SAFER System with \dtac{}s} \label{sec:safer}
\vspace{-7pt}


\begin{figure}[h] 
\vspace{-18pt}
\centering
\includegraphics[width=0.95\textwidth]{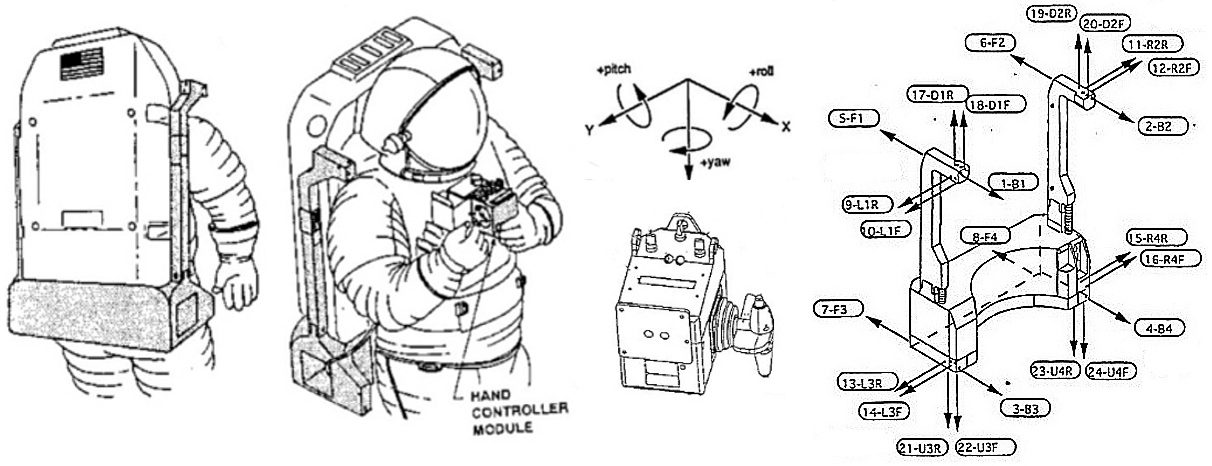}
\vspace{-10pt}
\caption{Left-to-right: SAFER deployed on an astronaut; the six degrees of movement axes (top); the hand controller (bottom); the mounting of the thrusters. (source: \cite{Kelly97})} \label{fig:safer}
\vspace{-12pt}
\end{figure}

\noindent Simplified Aid For Eva Rescue (SAFER) is a lightweight backpack propulsion system developed by NASA, and intended  to provide self-rescue capabilities for astronauts outside the spacecraft in space. It enabes movement along and around all three axes.
At frequent intervals, the software system reads commands from a \emph{hand-controller module} (HCM), and various sensors, and selects which of the mounted $24$ thrusters should be fired (by opening a vent releasing $\textit{GN}_2$ gas). In addition to the commands, the system contains an \emph{automatic attitude hold} (AAH), which attempts to nullify rotation. The system is shown in Fig. \ref{fig:safer}.

Several properties of a smaller version of the system have been verified using PVS \cite{Kelly97} and validated by testing using VDM-SL \cite{Agerholm97}. Here, a subset of this system is implemented in Dafny, and \dtac{}s are used to ensure freedom of \scode{null} references, together with a key
 functional property: only four thrusters can be simultaneously fired\footnote{Various 
snapshots from verifying these properties in Dafny 1.7.0  can be found at
\texttt{\url{https://sites.google.com/site/gudmundgrov/research/DafnySAFER.zip}}.}.
To encode this the following types are used: 
\begin{lstlisting}
datatype cmd      = NEG | ZERO | POS;
datatype thruster = B1 | B2 | B3 | B4 | F1 | ... ;
datatype switch   = TRAN | ROT;  
class T_CMD  { var X : cmd; var Y : cmd; var Z : cmd; ... }
class R_CMD  { var roll : cmd; var pitch : cmd; var yaw : cmd; ...}
class R_PRED { var roll : bool; var pitch : bool; var yaw : bool; ...}
class SD_CMD { var tran : T_CMD; var rot : R_CMD; ... }
\end{lstlisting}  
\scode{cmd} describes the orientation w.r.t the six axes, 
\scode{thruster} is the name of each thruster, while 
 \scode{switch} captures the mode of the HCM (see Fig.  \ref{fig:safer}). The class
\scode{T_CMD} holds the ``transitional" movements, \scode{R_CMD} ``rotational"
movement, and \scode{SD_CMD} combines them. \scode{R_PRED} is a predicate for each rotation axis. 

\lstset{language=dafny, frame=single, numbers = left,stepnumber=1,basicstyle=\scriptsize\sffamily} 
The HCM has a mode switch, and a stick that can be moved in four directions (along the three axes and twisting).  In each step, these values are read from the 
HCM and given as input to the \scode{control} method (note that \scode{|s|} is the length of sequence \scode{s}):
\begin{lstlisting}
method control(vert:cmd,horiz:cmd,trans:cmd,twist:cmd,mode:switch)  
 returns (thrusters : seq<thruster>) modifies this; { 
 ensures |thrusters| <= 4;  
    var aah_cmd := AAH_cmd();
    var grip_cmd := grip_command(vert,horiz,trans,twist,mode);
    var cmds := integrated_cmds (grip_cmd,aah_cmd);
    thrusters := selected_thrusters(cmds); }
\end{lstlisting}  
This method returns the set of thrusters to fire, and the key property is a poscondition of the method. The additional sensor readings are only required by the AAH. The four-thruster property is independent of the AAH, and has therefore been omitted from the program. Thus, we ignore the other sensors. However, we still need some inputs from the AAH, which are left undefined:
\lstset{language=dafny, numbers=none, frame=single,basicstyle=\scriptsize\sffamily}
\begin{lstlisting}
method AAH_cmd () returns (res : R_CMD) ensures res != null;
method AAH_ignore_HCM () returns (res : R_PRED) ensures res != null;
predicate method AAH_all_axis_off()
\end{lstlisting}
\scode{AAH_cmd} is the rotation command from the AAH. \scode{AAH_ignore_HCM}
is used in certain cases for the AAH to override rotational commands from the HCM. Finally, \scode{AAH_all_axis_off} is true if the AAH is off for all axes. Since these are left unimplemented, the property we prove is true regardless of what is returned from these methods. These three methods, together with \scode{control}, are the only `public' methods. The remaining, discussed next, are `private', meaning \dtac{}s are free to alter their contracts.

On line $4$ of \scode{control}, \scode{grip_command(vert,horiz,trans,twist,mode)} turns the inputs into a \scode{SD_CMD} object.
 If the mode is \scode{TRAN} then these are mainly transitional commands, otherwise they are mainly rotational commands. 
 Regardless of the mode, an \scode{X} and \scode{pitch} command will always be issued. On line $5$ of \scode{control}, the commands from 
the HCM and AAH are combined into a single command by 
\scode{integrated_cmds}.
Here, if all AAH axes are off, then 
 the command becomes the command from the HCM. Here, rotational commands have priority: if such are present,
 then only rotational commands are issued. If not, a transition command on one axis is issued,
 with the priority: $X > Y > Z$.
If it is not the case that all AAH axes are off, and
both AAH and HCM rotational commands
are present, then HCM has priority unless \scode{AAH_ignore_HCM} specifies that the HCM should be ignored for that
particular axis.

The key logic for the four thruster property is within the \scode{selected_thrusters} method, applied at the end of \scode{control}.
The encoding here follows directly from  \cite{Kelly97}. Firstly, the thrusters are divided  into those mounted in a Back or Front (BF) position, and those in a Left-Right-Up-Down (LRUD) position. Moreover, ``mandatory" and ``optional" thrusters are separated, creating four Dafny functions -- 
taking three axis commands and returning a sequence of thrusters. E.g.
\begin{lstlisting}
function method BF_optional(A:cmd,B:cmd,C:cmd) : seq<thruster> {
  match A case NEG => (match B case NEG => (match C case NEG => [B2,B3] ...
\end{lstlisting} 
returns the optional BF thrusters. Next, using these functions, methods for the 
BF and LRUD as a whole are created, exemplified for BF:
\begin{lstlisting}
method BF (A:cmd,B:cmd,C:cmd) returns (man:seq<thruster>,opt:seq<thruster>){  
 man := BF_mandatory(A,B,C); opt := BF_optional(A,B,C); }
\end{lstlisting} 
The \scode{selected_thrusters} method then combines these depending on whether or not \scode{X}, \scode{pitch} and \scode{yaw} commands are present:
\begin{lstlisting}
method selected_thrusters (comb : SD_CMD) returns (thrusters : seq<thruster>){
 var bf_main,bf_opt := BF(comb.tran.X,comb.rot.pitch,comb.rot.yaw);
 var lrud_main,lrud_opt := LRUD(comb.tran.Y,comb.tran.Z,comb.rot.roll);
 if (comb.tran.X == ZERO){
   if (comb.rot.pitch == ZERO && comb.rot.yaw == ZERO){
      thrusters := bf_opt + bf_main + lrud_opt + lrud_main;  
   } else{  thrusters := bf_opt + bf_main + lrud_main; }
 } else {
    if (comb.rot.pitch == ZERO && comb.rot.yaw == ZERO){
     thrusters := bf_main + lrud_opt + lrud_main; 
    }else{ thrusters := bf_main + lrud_main; }}}
\end{lstlisting} 

\vspace{-10pt}
\subsubsection{Handling null references} Dafny gives $16$ errors when trying
to verify the above example, and $15$ of them are ``target object may be null" thus
we need to rule out the possibility of \scode{null} references. We will illustrate 
a common strategy to prove such goals with one such failure in the
\scode{selected_thrusters} method. Here, the problem is the use of 
\scode{comb} in the call to \scode{BF}. First, the 
\dtname{null-to-assert} \dtac{} is applied, resulting in the following 
update to the code:
\begin{lstlisting}
 assert comb != null;
 var bf_main,bf_opt := BF(comb.tran.X,comb.rot.pitch,comb.rot.yaw);
\end{lstlisting} 
This assertion fails and the \dtname{assert-to-pre} \dtac{} is applied, 
which moves this assertion to a precondition:
\begin{lstlisting}
method selected_thrusters ... requires comb != null; ...
\end{lstlisting} 
Now this condition does not hold in the call to \scode{selected_thrusters}
by the \scode{control} method, raising the ``a precondition for this call might not hold" error. By applying the \dtname{pre-to-assert} \dtac{}, \scode{control} 
is updated as follows: 
\begin{lstlisting}
 var cmds := integrated_cmds (grip_cmd,aah_cmd,active_axis,ignore_HCM);
 assert cmds != null; thrusters := selected_thrusters(cmds); 
\end{lstlisting} 
This assertion fails to verify. We then apply \dtname{assert-to-post}
which moves this assertion to the postcondition of \scode{integrated_cmds}:
 \begin{lstlisting}
method integrated_cmds ... returns (comb : SD_CMD) ensures comb != null; 
\end{lstlisting}  
This completes the proof, and all such errors follow a similar approach. The only difference is that in certain cases, the final postcondition cannot be verified. In those cases the \dtname{post-to-assert} \dtac{}, which creates an assertion with the postcondition at the end of the branch,
is applied. This assertion is then moved upwards by \emph{assert-up}, and eventually becomes a precondition that is not met by the caller. This is
resolved by the \dtname{pre-to-assert} \dtac{}. At this point the above approach is followed. In total \dtname{null-to-assert} is applied 10 times; \dtname{assert-to-pre} 11 times; \dtname{pre-to-assert} 5 times;
\dtname{assert-to-post} 10 times; \dtname{assert-up} 19 times; and \dtname{post-to-assert} 4 times. 

\vspace{-10pt}
\subsubsection{Verifying the four thruster property}
After handling all the ``\scode{null} errors", the only open goal is the main property: the \scode{|thrusters| <= 4} postcondition of \scode{control}.
The main logic for this computation is within the 
\scode{selected_thrusters} method so the first step is to move the goal there:
\dtname{post-to-assert} moves it to an assertion in \scode{control}, and then \dtname{assert-to-post} moves it to a postcondition of 
\scode{selected_thrusters}\hspace{-3pt}.~
\dtname{post-to-assert} then moves the goal into the code, while 
\dtname{assert-up} (several) times, reduces this to 4 goals, one for each 
branch. 

First, the last branch is addressed. Here,  \dtname{assert-up}
moves the assertion ``over" the assignment, replacing \scode{thrusters}
with the assignment expression, resulting in:
\begin{lstlisting}
 assert |bf_main + lrud_main| <= 4; thrusters := bf_main + lrud_main; 
\end{lstlisting} 
\dtname{assert-up} is then applied until the assertion is just below the call to
\scode{BF}. The following rewrite rules will be required in the remainder:
\begin{small}
\vspace{-7pt}
\begin{eqnarray}
|~?x~ + ~?y~| \le ~?n & \longrightarrow & |~?x~| \le (?n~/~2) ~\wedge~ |~?y~| \le (?n~/~2) \label{rewr1} \\
|~?x~ + ~?y~| \le ~?n & \longrightarrow & |~?x~| ~\le~ ?n ~\wedge~ |~?y~| = 0 \label{rewr2}  \\
|~?x~ + ~?y~| \le ~?n & \longrightarrow & |~?x~| = 0 ~\wedge~ |~?y~| \le ~?n \label{rewr3} 
\end{eqnarray}
\vspace{-15pt}
\end{small}

\noindent We first apply \dtname{assert-rewr}((\ref{rewr1})) followed by \dtname{assert-conj-I}, resulting in:
\begin{lstlisting}
 assert |lrud_main| <= 4/2; assert |bf_main| <= 4/2; 
\end{lstlisting} 
This is followed by \dtname{assert-to-post}, and then \dtname{move-up}  followed by
\dtname{assert-to-post}, which moves these into a postcondition 
\scode{|man| <= 4/2} for both \scode{LRUD} and \scode{BF}, completing the first goal.

Next, we address the property at the first branch, by applying \emph{assert-up}:
\begin{lstlisting}
 assert |bf_opt + bf_main + lrud_opt + lrud_main| <= 4;
 thrusters := bf_opt + bf_main + lrud_opt + lrud_main;
\end{lstlisting} 
Then \dtname{assert-up-ctxt} twice, creates this assertion just after the call to \scode{LRUD}:
\begin{lstlisting}
assert comb.tran.X == ZERO && comb.rot.pitch == ZERO && comb.rot.yaw == ZERO 
                      ==> |bf_opt + bf_main + lrud_opt + lrud_main| <= 4;
\end{lstlisting}
A combination of \dtname{assert-rewr}((\ref{rewr1})), \dtname{assert-rewr}((\ref{rewr2})), \dtname{assert-rewr}((\ref{rewr3})) and \textit{assert-conj-I}
results in the following four assertions (with the same hypothesis as above):
\scode{|bf_opt| == 0}; \scode{|bf_main| == 0}; \scode{|lrud_opt| <= 4/2}; and
\scode{|lrud_main| <= 4/2}. \newline
\scode{|lrud_main| <= 4/2} is already handled by the postcondition of \scode{LRUD}
introduced by the first goal. Of the remaining, the first two goals are verified by \emph{move-up} followed by \emph{assert-to-post} for \scode{BF},
while the third one is verified by  \emph{assert-to-post} for \scode{LRUD}, after
\emph{assert-strengthen} has been applied trice to remove the conditions.

The same approach is applied to the third goal, where the condition is
\scode{comb.tran.X == ZERO} $\wedge$ $\lnot$\scode{(comb.rot.pitch == ZERO} $\wedge$ \scode{comb.rot.yaw == ZERO)}, and 
conditional assertions have the conclusions \scode{|bf_opt| == 0}, \scode{|bf_main| <= 2} and \newline \scode{|lrud_main| <= 2}.
Only the first of these is not provable, which becomes a postcondition for \scode{BF} by \emph{move-up} followed by \emph{assert-to-post}.

We follow the same approach for the final goal, which has the condition
\scode{comb.tran.X} $\neq$ \scode{ZERO} $\wedge$ \scode{comb.rot.pitch == ZERO} $\wedge$ \scode{comb.rot.yaw == ZERO}, and
the three goals \scode{|lrud_opt| == 0}, \scode{|bf_main| <= 2} and \scode{|lrud_main| <= 2}. In this case, only the first goal is not
proven. This property relies on the fact that there will only be one transition command at a time, thus, since $X$ is not \scode{ZERO},
that means that $Y$ and $Z$ are. We thus add the following precondition to \scode{selected_thrusters} 
\begin{lstlisting}
 requires comb.tran.X != ZERO ==> comb.tran.Y == ZERO && comb.tran.Z == ZERO;
\end{lstlisting}
by the \textit{pre-I} \dtac{}. As a result of this, the precondition to the \scode{selected_thrusters} call in \scode{control} is violated.
We then apply the \textit{pre-to-assert} \dtac{}, which adds the assertion before the call, followed by the \textit{assert-to-post} 
that adds this as a postcondition to \scode{integrated_commands}. This is then proven 
by a combination of \dtname{post-to-assert} and \dtname{assert-to-post} \dtac{}s to 
auxiliary methods (not given here).

The newly added precondition is moved into an assertion by \dtname{pre-to-assert}. 
Then, several \dtname{assert-down} \dtac{}s are applied so that it is moved 
just before the failing \scode{|lrud_opt| == 0}  assertion:
\begin{lstlisting}
 assert comb.tran.X != ZERO ==> comb.tran.Y == ZERO && comb.tran.Z == ZERO;
 assert comb.tran.X != ZERO ... ==> |lrud_opt| == 0
\end{lstlisting}
The \dtname{assert-comb1} \dtac{} combines these into
\begin{lstlisting}
 assert comb.tran.Y == ZERO && comb.tran.Z == ZERO ... ==> |lrud_opt| == 0
\end{lstlisting}
\dtname{assert-strengthen} (twice) removes \scode{pitch} and \scode{yaw} from the precondition (as these are not used by \scode{LRUD}):
\begin{lstlisting}
 assert comb.tran.Y == ZERO && comb.tran.Z == ZERO ==> |lrud_opt| == 0;
\end{lstlisting}
We then apply the \dtname{assert-to-post} for \scode{LRUD}, 
which completes the proof. In total, $57$ \dtac{} applications were required for the ``four thruster property".

%
%
%
%
%
%
%

\vspace{-5pt}
\section{Conclusion, Related \& Future Work}\label{sec:conc}
\vspace{-7pt}

By developing \dtac{}s and showing a detailed account of their usage to a  relevant case study
the hypothesis --  `\emph{it is possible to automate proofs by a notion of ``tactics" for program verifiers as a form of refactoring'}
-- has been validated.
However, understanding the degree of automation will require further examples, particularly those with iteration,
and importantly, implementation. As Dafny is open source, the implementation of \dtac{}s can exploit this existing  code base, 
and represent \dtac{}s as transformations on the 
abstract syntax tree. Such evaluation is likely to generalise, and hopefully improve, the \dtac{}s derived here, giving some indication of how to prune the search.
 The implementation of  SAFER  in Dafny was based upon previous embeddings in PVS  \cite{Kelly97} and 
VDM-SL \cite{Agerholm97}. SAFER has here been used as a proof of concept for
the feasibility of \dtac{}s, whilst \cite{Kelly97,Agerholm97} uses existing tools and approaches. A direct comparison between the approaches would therefore 
be unfair at this point. 

\dtac{}s are inspired by \emph{tactics} for ITP systems, originating from the 
LCF system \cite{GOR79a}. Most tactic languages are still based on this
paradigm.  To a less formal extent, refactoring also follows a ``trusted kernel" idea, where each refactoring should just make a small change to the code as this is easier to analyse \cite{Fowler99}. Larger refactorings then become compositions of smaller ones. \dtac{}s adopt the same approach, as most
\dtac{}s only make small changes, and the majority are special cases and/or compositions of smaller \dtac{}s.
Composition and specialisation are supported by sequential and branching combinators, and the use of arguments and partial instantiations. 
Other common combinators,  such as ``orelse", ``try" and ``repeat" are likely to be added in the future.


The proofs  conducted in the case-study and the \scode{LemmaLength} example are rather low-level and tedious at times. This is a consequence 
of working at the lowest ``kernel" level. Future work  would include the development of  higher level strategies which combine low-level rules.
Two examples illustrating that this should be possible have been seen: (1) the ``null reference" errors all followed the same strategy; and (2) three out of four of the ``four thruster" goals followed the same overall strategy. Moreover, the last goal only deviated by requiring an additional precondition. 


Chen \cite{Chen2010} describes a simple Dijkstra's guarded-command style imperative programming language, including both ``normal" execution statements as well as special verification statements. The verification statements can then be used to implement e.g. type checkers, abstract interpreters or contracts for the execution statements. This approach is much more general than DTacs, and whilst verification statements may be used to implement a similar notion of tactics, this is not discussed in \cite{Chen2010}. Moreover, it is not clear how such approach will work with an existing programming language and program verifier.

In \cite{Whiteside11}, refactoring of \emph{proof scripts} (for ITP systems) is addressed. Whilst it has similarities to \dtac{}s in the sense that both target proofs, it deviates as it is used to improve existing proofs -- albeit, it could be applied to generalise proofs into more general tactics. 
Further, \dtac{}s deviate from the case-based reasoning used in \cite{Monahan13} where specifications are carried between programs. 
\dtac{}s provides a mechanism for encoding generic strategies, which could be used to prove both the source and target programs in such a setting. 
A different approach to refactoring, is to ``optimise" the compilation into the IVL, as 
was done with the `induction tactic' for Dafny \cite{Leino12a}. 
It is important to note the difference between \dtac{}s and the vast number of static analysers such as abstract interpretation \cite{Cousot77}. There, the goal is to discover properties of programs, which could be utilised to prove properties.  \dtac{}s provide a way to enable users to encode their strategies when the proof is beyond such methods. However, in the future it would be interesting to see how such work could be utilised in a \dtac{}. Finally, cases of the \dtname{assert-up} \dtac{} bear resemblance to a \emph{predicate transformer} \cite{Dijkstra75} -- it should therefore be possible to encode other predicate transformers as \dtac{}s in the same way.
 
This paper  demonstrates that the use of ``tactics" for program verifiers is feasible and provides a firm foundation for further research in this area.


\vspace{-9pt}
\renewcommand{\refname}{}
\section*{References}
\vspace{-32pt}
\begin{small}
\bibliographystyle{plain}
\bibliography{../../dafny.bib}
\end{small}

\end{document}